

\magnification 1200

\hsize= 16 truecm
\vsize= 22 truecm

\hoffset=-0.0 truecm
\voffset= +1 truecm

\baselineskip=18 pt

\parindent 10pt 

\footline={\iftitlepage{\hfil}\else
       \hss\tenrm-- \folio\ --\hss\fi}	

\parskip 0pt 


\font\bello= cmr10 scaled \magstep3
\font\piccolo=cmr8

\def\eq{\autoeqno}
\def\re{\eqrefp}


\def\eq#1{\autoeqno{#1}}
\def\re#1{\eqrefp{#1}}

\newcount\notenumber \notenumber=1
\def\nota#1{\unskip\footnote{$^{\the\notenumber}$}{\piccolo #1}%
  \global\advance\notenumber by 1}

\def\s{\scriptstyle}  

\catcode`@=11 
%

\newcount\cit@num\global\cit@num=0

\newwrite\file@bibliografia
\newif\if@bibliografia
\@bibliografiafalse

\def\lp@cite{[}
\def\rp@cite{]}
\def\trap@cite#1{\lp@cite #1\rp@cite}
\def\lp@bibl{[}
\def\rp@bibl{]}
\def\trap@bibl#1{\lp@bibl #1\rp@bibl}

\def\refe@renza#1{\if@bibliografia\immediate        
    \write\file@bibliografia{
    \string\item{\trap@bibl{\cref{#1}}}\string
    \bibl@ref{#1}\string\bibl@skip}\fi}

\def\ref@ridefinita#1{\if@bibliografia\immediate\write\file@bibliografia{ 
    \string\item{?? \trap@bibl{\cref{#1}}} ??? tentativo di ridefinire la 
      citazione #1 !!! \string\bibl@skip}\fi}

\def\bibl@ref#1{\se@indefinito{@ref@#1}\immediate
    \write16{ ??? biblitem #1 indefinito !!!}\expandafter\xdef
    \csname @ref@#1\endcsname{ ??}\fi\csname @ref@#1\endcsname}

\def\c@label#1{\global\advance\cit@num by 1\xdef            
   \la@citazione{\the\cit@num}\expandafter
   \xdef\csname @c@#1\endcsname{\la@citazione}}

\def\bibl@skip{\vskip 0truept}


\def\stileincite#1#2{\global\def\lp@cite{#1}\global
    \def\rp@cite{#2}}
\def\stileinbibl#1#2{\global\def\lp@bibl{#1}\global
    \def\rp@bibl{#2}}

\def\citpreset#1{\global\cit@num=#1
    \immediate\write16{ !!! cit-preset = #1 }    }

\def\autobibliografia{\global\@bibliografiatrue\immediate
    \write16{ !!! Genera il file \jobname.BIB}\immediate
    \openout\file@bibliografia=\jobname.bib}

\def\cref#1{\se@indefinito                  
   {@c@#1}\c@label{#1}\refe@renza{#1}\fi\csname @c@#1\endcsname}

\def\cite#1{\trap@cite{\cref{#1}}}                  
\def\ccite#1#2{\trap@cite{\cref{#1},\cref{#2}}}     
\def\cccite#1#2#3{\trap@cite{\cref{#1},\cref{#2},\cref{#3}}}
\def\ccccite#1#2#3#4{\trap@cite{\cref{#1},\cref{#2},\cref{#3},\cref{#4}}}
\def\ncite#1#2{\trap@cite{\cref{#1}--\cref{#2}}}    
\def\upcite#1{$^{\,\trap@cite{\cref{#1}}}$}               
\def\upccite#1#2{$^{\,\trap@cite{\cref{#1},\cref{#2}}}$}  
\def\upncite#1#2{$^{\,\trap@cite{\cref{#1}-\cref{#2}}}$}  

\def\clabel#1{\se@indefinito{@c@#1}\c@label           
    {#1}\refe@renza{#1}\else\c@label{#1}\ref@ridefinita{#1}\fi}

\def\biblskip#1{\def\bibl@skip{\vskip #1}}           

\def\insertbibliografia{\if@bibliografia             
    \immediate\write\file@bibliografia{ }
    \immediate\closeout\file@bibliografia
    \catcode`@=11\input\jobname.bib\catcode`@=12\fi}


\def\commento#1{\relax} 
\def\biblitem#1#2\par{\expandafter\xdef\csname @ref@#1\endcsname{#2}}



%
%
\def\b@lank{ }


\newif\iftitlepage      \titlepagetrue

\def\titoli#1{
         \xdef\prima@riga{#1}\voffset+20pt
        \headline={\ifnum\pageno=1
             {\hfil}\else\hfil{\piccolo \prima@riga}\hfil\fi}}

\def\duetitoli#1#2{
                    \voffset=+20pt
                    \headline={\iftitlepage{\hfil}\else
                              {\ifodd\pageno\hfil{\piccolo #2}\hfil
             \else\hfil{\piccolo #1}\hfil\fi}\fi} }

\def\la@sezionecorrente{0}

\catcode`@=12

%
\catcode`@=11 
%
%
\def\b@lank{ }

\newif\if@simboli
\newif\if@riferimenti
\newif\if@bozze

\newwrite\file@simboli
\def\simboli{
    \immediate\write16{ !!! Genera il file \jobname.SMB }
    \@simbolitrue\immediate\openout\file@simboli=\jobname.smb}

\def\bozze{\@bozzetrue}

\newcount\eq@num\global\eq@num=0
\newcount\sect@num\global\sect@num=0

\newif\if@ndoppia
\def\numerazionedoppia{\@ndoppiatrue\gdef\la@sezionecorrente{\the\sect@num}}

\def\se@indefinito#1{\expandafter\ifx\csname#1\endcsname\relax}
\def\spo@glia#1>{} 

\newif\if@primasezione
\@primasezionetrue

\def\s@ection#1\par{\immediate
    \write16{#1}\if@primasezione\global\@primasezionefalse\else\goodbreak
    \vskip\spaziosoprasez\fi\noindent
    {\bf#1}\nobreak\vskip\spaziosottosez\nobreak\noindent}
%

\def\sezpreset#1{\global\sect@num=#1
    \immediate\write16{ !!! sez-preset = #1 }   }

\def\spaziosoprasez{26pt plus5pt minus3pt}
\def\spaziosottosez{15pt}

\def\sref#1{\se@indefinito{@s@#1}\immediate\write16{ ??? \string\sref{#1}
    non definita !!!}
    \expandafter\xdef\csname @s@#1\endcsname{??}\fi\csname @s@#1\endcsname}

\def\autosez#1#2\par{
    \global\advance\sect@num by 1\if@ndoppia\global\eq@num=0\fi
    \xdef\la@sezionecorrente{\the\sect@num}
    \def\usa@getta{1}\se@indefinito{@s@#1}\def\usa@getta{2}\fi
    \expandafter\ifx\csname @s@#1\endcsname\la@sezionecorrente\def
    \usa@getta{2}\fi
    \ifodd\usa@getta\immediate\write16
      { ??? possibili riferimenti errati a \string\sref{#1} !!!}\fi
    \expandafter\xdef\csname @s@#1\endcsname{\la@sezionecorrente}
    \immediate\write16{\la@sezionecorrente. #2}
    \if@simboli
      \immediate\write\file@simboli{ }\immediate\write\file@simboli{ }
      \immediate\write\file@simboli{  Sezione 
                                  \la@sezionecorrente :   sref.   #1}
      \immediate\write\file@simboli{ } \fi
    \if@riferimenti
      \immediate\write\file@ausiliario{\string\expandafter\string\edef
      \string\csname\b@lank @s@#1\string\endcsname{\la@sezionecorrente}}\fi
    \goodbreak\vskip 48pt plus 60pt
    \noindent\if@bozze\llap{\it#1\quad }\fi
      {\bf\the\sect@num.\quad #2}\par\nobreak\vskip 15pt
    \nobreak\noindent}

\def\semiautosez#1#2\par{
    \gdef\la@sezionecorrente{#1}\if@ndoppia\global\eq@num=0\fi
    \if@simboli
      \immediate\write\file@simboli{ }\immediate\write\file@simboli{ }
      \immediate\write\file@simboli{  Sezione ** : sref.
          \expandafter\spo@glia\meaning\la@sezionecorrente}
      \immediate\write\file@simboli{ }\fi
    \s@ection#2\par}


\def\eqpreset#1{\global\eq@num=#1
     \immediate\write16{ !!! eq-preset = #1 }     }

\def\eqref#1{\se@indefinito{@eq@#1}
    \immediate\write16{ ??? \string\eqref{#1} non definita !!!}
    \expandafter\xdef\csname @eq@#1\endcsname{??}
    \fi\csname @eq@#1\endcsname}

\def\eqlabel#1{\global\advance\eq@num by 1
    \if@ndoppia\xdef\il@numero{\la@sezionecorrente.\the\eq@num}
       \else\xdef\il@numero{\the\eq@num}\fi
    \def\usa@getta{1}\se@indefinito{@eq@#1}\def\usa@getta{2}\fi
    \expandafter\ifx\csname @eq@#1\endcsname\il@numero\def\usa@getta{2}\fi
    \ifodd\usa@getta\immediate\write16
       { ??? possibili riferimenti errati a \string\eqref{#1} !!!}\fi
    \expandafter\xdef\csname @eq@#1\endcsname{\il@numero}
    \if@ndoppia
       \def\usa@getta{\expandafter\spo@glia\meaning
       \la@sezionecorrente.\the\eq@num}
       \else\def\usa@getta{\the\eq@num}\fi
    \if@simboli
       \immediate\write\file@simboli{  Equazione 
            \usa@getta :  eqref.   #1}\fi
    \if@riferimenti
       \immediate\write\file@ausiliario{\string\expandafter\string\edef
       \string\csname\b@lank @eq@#1\string\endcsname{\usa@getta}}\fi}

\def\autoreqno#1{\eqlabel{#1}\eqno(\csname @eq@#1\endcsname)
       \if@bozze\rlap{\it\quad #1}\fi}
\def\autoleqno#1{\eqlabel{#1}\leqno\if@bozze\llap{\it#1\quad}
       \fi(\csname @eq@#1\endcsname)}
\def\eqrefp#1{(\eqref{#1})}
\def\numeriadestra{\let\autoeqno=\autoreqno}
\def\numeriasinistra{\let\autoeqno=\autoleqno}
\numeriadestra

\catcode`@=12

\numerazionedoppia


\autobibliografia


\def\bbuildrel#1_#2{\mathrel{
\mathop{\kern 0pt#1}\limits_{#2}}}

\titlepagetrue

\vglue 1 truecm

\centerline {\bello    Structure of metastable states in spin glasses  }
\centerline {\bello         by means of a three replica potential    }

\vskip 1 truecm

\centerline {Andrea Cavagna, Irene Giardina, Giorgio Parisi}

\vskip 1 truecm

\centerline {\it          Dipartimento di Fisica}
\centerline {\it  Universit\`a di Roma I, "La Sapienza"}
\centerline {\it   P.le A. Moro 5, 00185 Roma, Italy }
\centerline {\it              and                    } 
\centerline {\it INFN Sezione di Roma I, Roma, Italy.}
\vskip 0.5 truecm

\centerline {\sl cavagna@roma1.infn.it}
\centerline {\sl giardina@roma1.infn.it}
\centerline {\sl parisi@roma1.infn.it}

\vskip 0.5 truecm

\centerline{November 7, 1996}

\vskip 2 truecm

\centerline{\bf Abstract}

\vskip 0.5 truecm

{\piccolo \noindent 
In this paper we introduce a three replica potential useful to examine
the structure of metastable states above the static 
transition temperature $\s T_c$, in the spherical $\s p$-spin model. 
Studying the minima of the potential we are able to find which is the 
distance between the nearest equilibrium and local equilibrium 
states, obtaining in this way information on the 
dynamics of the system. Furthermore, the analysis of the potential at 
the dynamical transition temperature $\s T_d$ suggests that equilibrium 
states are not randomly distributed in the phase space.}

\vfill\eject


\titlepagefalse
\autosez{I} Introduction.
\par

In these years it has been realized that there are model systems for which the 
dynamics can be analytically studied in the infinite volume limit.  In some of 
these systems, below a characteristic temperature $T_d$, called dynamical
transition temperature,
the phase space of the 
equilibrium configurations breaks down in regions (from here on valleys) where 
the system is trapped for an infinite time.  Therefore, in this situation, 
if the 
system is near an equilibrium configuration at the initial time, it will 
remain in this region forever.  The number of these regions is exponentially 
high as soon as $T>T_c$, where $T_c<T_d$ is the static transition temperature, 
at which only a 
finite number of valley starts to dominate the partition function; these
features are clearly revealed by an analysis based on the TAP equations 
\cccite{tap}{kpz}{crisatap}.  
All the equilibrium valleys are characterized by the same energy density $E$,
entropy density $S_v$ and self-overlap $q_{EA}$. 
The total entropy density of the system will be given by
$$
S= S_v +\Sigma \ ,
\eq{uno}
$$
where $\Sigma$ is the so called configurational entropy or complexity, i.e. 
$\Sigma=(1/N)\ln{\cal N}$,
where $\cal N$ is the exponentially high number of valleys and $N$ is the
size of the system. 
The complexity goes to zero at the transition 
temperature $T_c$.

It is also 
true that in the 
region $T<T_d$, if the system starts randomly at the initial time, the energy of 
the system evolves toward a value greater than the equilibrium one 
\ccite{kpz}{ck1}.  In other 
words, metastable states are present. In the rest of the paper these metastable 
states will be also called local equilibrium state, while the name 
equilibrium states will be reserved to valleys which have the correct equilibrium 
energy density.

It is clear that these features, especially the last one, are likely to be an 
artefact of the mean field approximation, which is correct in these models. It 
is therefore interesting to try to understand what happens when  corrections 
to the mean field theory are present. The simplest case we can think about is 
when $N$ is finite and large: the times to escape from a valley are likely to 
be exponentially high in $N$. The direct computation of these exponentially 
large times using the dynamical equations is rather involved and it has not 
been done. A partial solution to the problem consists in assuming the dynamics 
is do\-mi\-na\-ted at ultra large times by the crossing of free energy 
barriers between different valleys and some information can thus be 
gathered by computing these barriers and by the analysis of the
mutual disposition of metastable states.

This program cannot be pursued directly with 
the usual tools of statistical mechanics and we must resort to use a different 
method, the {\it real replica method} \ccite{kpz}{franzparisi}, in which two 
or more (in general $M$) real 
replicas are introduced.  If we denote by $\bf q $ the $M \times M$ matrix of 
the overlaps imposed among the different real replicas
$$
q_{a,b}= {1\over N} \sum_i \sigma^a_i \sigma^b_i \ , 
\eq{due}
$$
detailed information on the free energy landscape is carried by the free 
energy as function of  ${\bf q}$.
This program is quite recent. Up to this moment only the case $M=2$ has been 
examinated in details. In this case one computes a two replica potential
$V_2(q_{12})$, which is the free energy of replica 2 constrained to have overlap 
$q_{12}$ with replica 1, supposed to stay at equilibrium 
\cite{franzparisi}. 
A different potential $W_{12}(q_{12})$ can be introduced, 
defined as  the 
free energy increase of  a pair of two replicas if we constrain them to 
stay at mutual overlap $q_{12}$ \cite{kpz}. 
The main difference among these two cases is the following: 

\noindent
In the first case  replica 1 is at equilibrium and  replica 2 is not,
because it must satisfy a constraint; obviously for some 
particular values of $q_{12}$ (e.g. $q_{12}=0$) the constraint is harmless and 
also replica 2 is at equilibrium.
In the second case replica 1 and 2 are chosen in such a way to satisfy a 
constraint and both will be in general out of equilibrium.
The first construction is the most appropriate one if we want to get 
information on the free energy landscape around an equilibrium 
configuration. The relevant 
computations have been already done and they will be summarized here for reader 
convenience. From this two replica potential one can compute the dynamical 
transition and the configurational entropy, and obtain a first estimate of the 
barriers separating different valleys.

The question we address in this paper is the organization of equilibrium and 
local equilibrium states.  For example, given a generic equilibrium 
configuration we would like to know which is the maximum overlap $\bar q$ 
at which is found a local equilibrium state. In the same 
fashion we would like to know which is the maximum overlap $q^\star$ 
at which is found an equilibrium state. It is not evident {\it a priori}
if $\bar q\ne q^\star$ (which is the result we find) or not.

In order to answer to this and other questions we need to consider a three 
replica potential
$V_3(q_{12},q_{13},q_{23})$,
which is the free energy of  replica 3 constrained to stay at overlap $q_{13}$ 
and $q_{23}$ from replica 1 and replica 2 respectively, where replica 1 is an 
equilibrium configuration and replica $2$ is constrained to stay at distance 
$q_{12}$ from replica 1. This potential is useful to explore in more detail the 
free energy landscape with respect to the previous case where only two replicas 
are present. The stationary points of this potential correspond to local 
equilibrium states of the system.

In this paper we devote our attention to the $p$-spin spherical model, due to 
its simplicity and to the relatively simple form of the free energy.  Similar 
considerations can be applied also to other models, but this will not be done in 
this paper.  In Sect. 2 we recall for reader convenience the definition of the 
$p$-spin spherical model and the computation of the two replica potential.  In 
Sect. 3 we present the computation for the three replica potential.  The shape of 
the potential and its stationary points, which correspond to local equilibrium 
states, are computed in Sect. 4. In Sect. 5 we study the temperature 
dependence of the various quantities involved. Finally, in Sect. 6 we discuss 
some of the implications of our finding for the dynamics of the system at large 
volume. Two appendices contain the more technical aspects of the 
computations.
\vfill\eject


\autosez{v2} The two replica potential.
\par

Let us introduce the $p$-spin spherical model 
\ccccite{grome}{gard}{tirumma}{crisa1}, defined by the Hamiltonian
$$
\eqalign{
H(\sigma)= & \sum_{i_1<i_2<\dots<i_p} J_{i_1\dots i_p} 
\sigma_{i_1}\dots \sigma_{i_p}      \cr
& {1\over N}  \sum_i \sigma_i^2=  1   \ , \cr}
\eq{model}
$$
where the $\sigma_i$ are real variables satisfying the spherical constraint;
the couplings $J_{i_1\dots i_p}$ are independent Gaussian variables
with variance
$$
\overline{J_{i_1\dots i_p}^2}={p!\over 2 N^{p-1}}   \ .
\eq{varu}
$$
With these definitions
$$
\overline{ H(\sigma) H(\sigma') }\buildrel\hbox{\sevenrm def}\over=
N{1\over 2} f(q_{\sigma\sigma'})=
N{1\over 2} q^p_{\sigma\sigma'} \ ,
\eq{zz}
$$
where $q_{\sigma\sigma'}$ is the overlap between the two configurations 
(see eq. \re{due}).
We note that it is possible a generalization to random Hamiltonian models, 
specified by different forms of the correlation
function $f$ \ccite{mezpa}{nieu}.

Following \cite{franzparisi} we consider the  two replica potential
$V_2(q_{12})$, defined as the free energy cost to keep a configuration 
$\tau$ (replica 2) at a fixed overlap $q_{12}$ with an equilibrium
configuration $\sigma$ (replica 1):
$$
-\beta N V_2(q_{12})= -\beta N (F_{fixed} - F_{free})=  
$$
$$
\overline{
{1\over Z_{free}}\int\ d\sigma \exp(-\beta H(\sigma))
\ \log\left(\int\ d\tau \exp(-\beta H(\tau))\ 
\delta(q_{\sigma\tau}-q_{12})\right)}
- \overline{\log Z_{free}} \     .
\eq{init}
$$
It is important to note that in this way we compute the free energy 
only 
of the second replica $\tau$, while we operate an annealed average over 
$\sigma$. Indeed, this can be done
by virtue of the assumption that the constrained free energy 
$F_{fixed}$ is self-averaging with respect to $\sigma$. 
To perform the average on the quenched disorder, we use the replica trick:
\eject
$$
-\beta N V_2(q_{12})=
$$
$$
\eqalign{
=\lim_{n,m\rightarrow 0}{1\over m} \log & \overline{
\int\ d\sigma_a d\tau_b 
\exp\left[- \beta\left(\sum_{a=1}^n H(\sigma_a)+
\sum_{b=1}^m H(\tau_b)\right)\right] 
\prod_{b=1}^m \delta(q_{\sigma_1\tau_b}-q_{12})}          \cr
& \phantom{    \int\ d\sigma_a d\tau_b 
               \exp- \beta\sum_{a=1}^n 
               H(\sigma_a)+\sum_{b=1}^m H(\tau_b)     }
-\overline{ \log Z_{free}  }   \ .                \cr}
\eq{replica}
$$
We introduce the overlap sub-matrices
$$
\eqalign{
Q^{11}_{ab}=&\ q_{\sigma_a\sigma_b}  \quad \quad a=1,\dots,n; \  b=1,\dots,n
\cr
Q^{12}_{ab}=&\ q_{\sigma_a\tau_b}    \quad \quad a=1,\dots,n; \  b=1,\dots,m
\cr
Q^{22}_{ab}=&\ q_{\tau_a\tau_b}     \quad \quad a=1,\dots,m; \  b=1,\dots,m \ ;
\cr}
\eq{blo}
$$
and the total matrix of the overlap  $\bf Q$, formed by the blocks 
\re{blo}. With this definition,  equation \re{replica} gives
$$
-\beta V_2(q_{12})=
\lim_{n,m\rightarrow 0}\ {1\over 2m} 
\left( \beta^2 \sum_{a,b} f({\bf Q}_{ab})+
\log \det {\bf Q} \right) + \beta  F_{free} \ .
\eq{snaz}
$$

Before proceeding with the calculation, it is necessary to give sensible
ansatz for the overlap sub-matrices. From now on we will consider only temperatures $T$ greater than $T_c$ and zero external magnetic field; therefore, since
 $Q^{11}$ is referred to an independent free system,  we 
assume for it the symmetric form
$$
Q^{11}_{ab}=\delta_{ab}   \ .
\eq{q11}
$$
The structure of $Q^{12}$ is in part imposed by the constraint in 
\re{replica}, that forces the first row to be equal to $q_{12}$. 
The simplest ansatz for the whole matrix is 
$$
Q^{12}_{ab}=q_{12}\ \delta_{a,1}  \ .
\eq{q12}
$$
This choice is confirmed by the analysis of the potential 
in the limit 
\nota{
At $\s \beta= 0$ the calculation reduces
to a purely geometrical problem; indeed, in this case the potential is 
simply proportional to the volume of the phase space region
accessible to the constrained replica.
Therefore, the comparison with this limit is crucial to test the ansatz
for the overlap matrices.
}
\ $\beta\rightarrow 0$; moreover, if we introduce an additional 
parameter $w_{12}$ for the rest of the 
matrix, the saddle point equations lead to the solution $w_{12}=0$.

In \cite{franzparisi} $Q^{22}$ was assumed symmetric, but a successive 
examination suggested that a more general one step RSB 
form has to be taken \cite{franz}, with variational parameters $(x_r,r_1,r_0)$:
$$
Q^{22}_{ab}=(1-r_1)\delta_{ab}+(r_1-r_0)\epsilon_{ab}+r_0 \ ,
\eq{qw22}
$$
where $\epsilon_{ab}$ is equal to one in the diagonal blocks of size $x_r$ and zero
elsewhere.

With these assumptions we obtain (see Appendix A for details):
$$
\eqalign{
-2\beta V_2(q_{12})= &               
\ 2\beta^2 f(q_{12})+
\beta^2 (x_r-1)f(r_1)-\beta^2 x_r f(r_0)     \cr
+ & \log(1-r_1)+{1\over x_r}\log\left(
1+x_r{r_1-r_0\over 1-r_1}\right)+
{r_0-q_{12}^2\over 1-r_1+x_r(r_1-r_0)}    \ .  \cr}
\eq{v2}
$$
The parameters with respect to which the potential has to be maximized
are $(x_r,r_1,r_0)$ and the saddle point equations read:
$$
\eqalign{
(1-x_r)\beta^2 f'(r_1)= & \ (1-x_r)\ {1\over (1-r_1+x_r(r_1-r_0))^2}
\left( r_1-q_{12}^2+x_r {(r_1-r_0)^2\over 1-r_1}\right)          \cr
x_r\beta^2 f'(r_0)= & \ x_r\ {(r_0-q_{12}^2)\over (1-r_1+x_r(r_1-r_0))^2}  \cr
\beta^2 f(r_1)-\beta^2 f(r_0)= & \ {1\over x_r^2} \log\left(
1+x_r{r_1-r_0\over 1-r_1}\right)         \cr
&\phantom{{1\over x_r^2} \log}    -{1\over x_r}(r_1-r_0)
\ {1-r_1+x_r(r_1-2r_0+q_{12}^2)\over(1-r_1+x_r(r_1-r_0))^2}   \  .\cr }
\eq{v2sel}
$$
Setting $r_0=r_1$ we obviously recover the replica symmetric expressions 
given in \cite{franzparisi}.

The RSB ansatz for $Q^{22}$ slightly modifies the shape of the potential $V_2$ 
(see Fig. 1)
eliminating the secondary minimum found in \cite{franzparisi}, the 
interpretation of which was unclear.  For the rest, all the features of the 
potential remain the same: there is an absolute minimum for $q_{12}=0$ with 
$V_2=0$, which represents a typical equilibrium configuration of the second 
replica $\tau$; moreover for $T<T_d$ there is a relative minimum $M$ for 
$q_{12}=q_{EA}$ corresponding to the situation in which $\tau$ is in the same 
pure state as the first replica $\sigma$.  We note that in this second 
situation, the configuration $\tau$ is of equilibrium but not typical, in the 
sense that, for entropic reasons, if the system were not constrained, it would 
be very unlikely to find $\tau$ in the same pure state of $\sigma$.  It turns 
out that the value of the potential in $M$ gives information on the number of 
dominant pure states, i.e.
$$ V_2(q_{EA})=T \Sigma \ ,
\eq{sigma}
$$
where $\Sigma$ is the configurational entropy \cite{franzparisi}.

\includegraphics{snacio_fig1.ps}

\vbox{
      \hbox{     \vbox{$\s V_2$ \vglue 2.7 truecm}
                 \vbox{\vglue 6.5 truecm}                             }
      \hbox{     \hglue 5.5 truecm $\s q_{12}$                        }
      \hbox{     \piccolo{\hskip 1 truecm Fig. 1: The potential 
                 $\s V_2$ at $\s \beta=1.64$ and $\s p=3$; 
                 $\s q_{EA}=0.546$.}                                  }
                                                                           }

\autosez{v3} The three replica potential.
\par

As we have seen, the analysis of $V_2$ gives information on
the self-overlap and on the number  of the states dominating at a certain
temperature. It remains rather unclear how the phase space is organized 
in terms of these states; in other words, we would like to know more
about their mutual overlaps, and study the evolution of the whole structure
with the temperature. 

A natural generalization of the method that leads to $V_2$, consists in 
considering three different real replicas, constrained to have fixed
mutual overlap, as we are going to explain.
As in the case of $V_2$, we take a first independent replica $\rho$ which
thermalizes freely at a temperature $T$. We then consider a second replica
$\sigma$ constrained to thermalize at a fixed overlap $q_{12}$ with $\rho$.
Finally, we calculate the free energy of a third replica $\tau$ constrained
to have overlap $q_{13}$ with $\rho$ and $q_{23}$ with $\sigma$.
In this way it is possible to define a three replica potential $V_3$, taking
the difference between the constrained free energy of $\tau$ and the
free energy of the unconstrained system:
$$   
-\beta N V_3(q_{12},q_{13},q_{23})= 
$$
$$
\eqalign{ 
= & \overline{   {1\over Z_{free}}\int d\rho\exp(-\beta H(\rho))
{1\over Z(\rho;q_{12})}\int d\sigma\exp(-\beta H(\sigma))
\ \delta(q_{\rho\sigma}-q_{12}) }   \times                \cr
& \phantom{provare} \overline{
\log\int d\tau\exp(-\beta H(\tau))\ \delta(q_{\rho\tau}-q_{13})
\ \delta(q_{\sigma\tau}-q_{23})}  
-\overline{\log Z_{free} }   \ , \cr}
\eq{v3}  
$$
with
$$
Z(\rho;q_{12})=\int d\sigma\exp(-\beta H(\sigma))
\ \delta(q_{\rho\sigma}-q_{12})  \ .
\eq{zetaro}
$$
In our analysis we limit ourselves to the situation in which the temperatures
are the same for the three replicas.
Our purpose is to study the potential $V_3$ in the plane $(q_{13},q_{23})$ 
at various fixed values of $q_{12}$. In this way, through the position of 
replica 3, we can explore the neighborhood of 
replica 2, which is not a configuration  of equilibrium 
for the free system; this is possible because of the different role of 
the first and second replica.
By the usual trick we get
$$   
-\beta N V_3(q_{12},q_{13},q_{23})= 
$$
$$
\eqalign{
= & \lim_{n,m,l\rightarrow 0} {1\over l}\log  \overline{
\int d\rho_a d\sigma_b d\tau_c\exp\left[-\beta\left(
\sum_{a=1}^n H(\rho_a)
+\sum_{b=1}^m H(\sigma_b)
+\sum_{c=1}^l H(\tau_c)\right)\right]            } \times                  \cr
& \phantom{provare} \overline{ \prod_{b=1}^m\ \delta(q_{\rho_1\sigma_b}-q_{12})
\prod_{c=1}^l\ \delta(q_{\rho_1\tau_c}-q_{13})\delta(q_{\sigma_1\tau_c}-q_{23})}
-\overline{\log Z_{free}    }        \    .                              \cr}
\eq{v3rep}
$$
Again we introduce the sub-matrices:
$$
\eqalign{
Q^{11}_{ab}=&\ q_{\rho_a\rho_b}  \quad \quad a=1,\dots,n; \  b=1,\dots,n
\cr
Q^{12}_{ab}=&\ q_{\rho_a\sigma_b}    \quad \quad a=1,\dots,n; \  b=1,\dots,m
\cr
Q^{22}_{ab}=&\ q_{\sigma_a\sigma_b}     \quad \quad a=1,\dots,m; \  b=1,\dots,m 
\cr
Q^{13}_{ab}=&\ q_{\rho_a\tau_b}    \quad \quad a=1,\dots,n; \  b=1,\dots,l
\cr
Q^{23}_{ab}=&\ q_{\sigma_a\tau_b}    \quad \quad a=1,\dots,m; \  b=1,\dots,l
\cr
Q^{33}_{ab}=&\ q_{\tau_a\tau_b}    \quad \quad a=1,\dots,l; \  b=1,\dots,l \ ;
\cr}
\eq{cucu}
$$
and the total overlap matrix $\bf Q$.
In analogy with the case of $V_2$, the expression of $V_3$ in terms of $\bf Q$
after averaging  over the disorder, is
$$
-\beta V_3(q_{12},q_{13},q_{23})=
\lim_{n,m,l\rightarrow 0}\ {1\over 2l} 
\left( \beta^2 \sum_{a,b} f({\bf Q}_{ab})+
\log \det {\bf Q} \right) + \beta  F_{free} \ .
\eq{tuka}
$$
It is important to bear in mind the difference between the roles of replicas 1 
and 2, which is
encoded in the different form of the relative overlap matrices.
$Q^{11}$ refers to the first free replica, while $Q^{22}$ refers to replica 2,
which thermalizes at {\it fixed} overlap $q_{12}$ with 1. Both replicas are
however independent from the third one, so the form of $Q^{11}$, $Q^{12}$ and
$Q^{22}$ is identical to the one assumed in the case of $V_2$ and the value 
of the variational parameters $(x_r,r_1,r_0)$, which are function of $q_{12}$, 
are determined by the same equations \re{v2sel}. 

In second place, $Q^{33}$ is assumed one step RSB, with parameters 
$(x_s,s_1,s_0)$:
$$
Q^{33}_{ab}=(1-s_1)\delta_{ab}+(s_1-s_0)\epsilon_{ab}+s_0  \  ;
\eq{qq33}
$$
there is no reason to set $x_s=x_r$, since just as $Q^{22}$ is broken 
independently from $Q^{11}$, the breakings of $Q^{22}$ and $Q^{33}$ have to be
assumed independent; moreover, there are no algebraic motivations for this
equality.

For what concerns $Q^{13}$, the limit $\beta\rightarrow0$ shows 
that 
it is correct to assume a form similar to $Q^{12}$,
$$
Q^{13}_{ab}=q_{13}\ \delta_{a,1}  \ .
\eq{qq13}
$$ 

However, the same limit shows that this form is no longer valid for $Q^{23}$;
namely, the first row will still be equal to $q_{23}$, but the rest of the 
matrix no more can be set equal to zero. 
Moreover, the structure of
$Q^{23}$ has to take into account the RSB form of $Q^{22}$: if
replicas 2 are organized into clusters, it can be that the overlap
of 3 with the particular cluster that contains the {\it first} 
replica $\sigma_1$ of 2 
(this is the only replica of 2 {\it really} coupled to 3, as shown in 
\re{v3rep})
is different from the overlap of 3 with any other cluster. 

Therefore, we postulate a RSB form also for 
$Q^{23}$. This is possible if we break the symmetry in the rows of 
$Q^{23}$ in the following way:
$$
Q^{23}_{ab}=(q_{23}-w_{23})\delta_{a,1}+(w_{23}-z_{23})\epsilon_{a,1}+z_{23} \ .
\eq{qq23}
$$
This one step RSB needs a breaking point $x$, but the genesis of the
breaking shows clearly that for $Q^{23}$ it has to be taken as breaking
point $x_r$, that is the same of $Q^{22}$.

In conclusion, the new variational parameters are:
$$
x_s, s_1, s_0, w_{23}, z_{23} \  .
\eq{var}
$$

With these ansatz, the expression of $V_3$ is (see Appendix B for details)
$$
\eqalign{
-2\beta V_3(q_{12},q_{13},q_{23})= &
\ 2\beta^2 f(q_{13})+2\beta^2 f(q_{23})+2\beta^2(x_r-1) f(w_{23})-
2\beta^2 x_r f(z_{23})                                \cr
+ & \ \beta^2 (x_s-1) f(s_1)-\beta^2 x_s f(s_0)         \cr
+ & \log(1-s_1)+{1\over x_s}\log\left(
1+x_s{s_1-s_0\over 1-s_1}\right)+
{s_0-y\over 1-s_1+x_s(s_1-s_0)}            \ ,           \cr}
\eq{v3fin}
$$
with
$$
\eqalign{
y=q_{13}^2+ & (q_{23}-w_{23})^2 (a+b+c)+2(q_{23}-w_{23})(w_{23}-z_{23})
(a+x_r(b+c))    \cr
          + & 2(q_{23}-w_{23})(z_{23}-q_{12}q_{13})(a+b x_r)+
(w_{23}-z_{23})^2 x_r (a+x_r(b+c))          \cr
          + & 2(w_{23}-z_{23})(z_{23}-q_{12}q_{13}) x_r (a+b x_r)  \ ,\cr}
\eq{y2}
$$
and
$$
\eqalign{
a= & \ {1\over 1-r_1}    \cr
b= & - {r_1-r_0 \over 1-r_1 }\cdot {1\over 1-r_1 +x_r(r_1-r_0)}   \cr
c= & - {r_0-q_{12}^2\over (1-r_1 +x_r(r_1-r_0))^2}  \  .       \cr  }
\eq{cba}
$$

Expression \re{v3fin} is very similar to the one of $V_2(q_{12})$ with 
$(x_s,s_1,s_0)$ playing the role of $(x_r,r_1,r_0)$ and $y$ instead of
$q_{12}^2$. Indeed, the saddle point equations for $(x_s,s_1,s_0)$ are 
in form exactly identical to \re{v2sel}, changing $q_{12}^2$ into $y$.  
Moreover, $V_3$ has to be maximized with respect to $w_{23}$ and $z_{23}$:
$$
\eqalign{
(1-x_r)\beta^2 f'(w_{23})= & \ (1-x_r)\ {1\over 1-s_1+x_s(s_1-s_0)}
[(q_{23}-w_{23})(b+c)   \cr
          & \phantom{(1-x_r)}+(w_{23}-z_{23})(a+x_r(b+c))+(z_{23}-q_{12}q_{13})
(a+b x_r)]         \cr
x_r\beta^2 f'(z_{23})= & \ x_r\ {1\over 1-s_1+x_s(s_1-s_0)}
[  (q_{23}-w_{23})c   \cr
          & \phantom{x_r} +x_r(w_{23}-z_{23})c+(z_{23}-q_{12}q_{13})(a+b x_r)] 
    \   .  \cr   }
\eq{xz}
$$ 
We note that the equations for $(x_s,s_1,s_0)$ are coupled to equations 
\re{xz} by means of $y$, that contains $w_{23}$ and $z_{23}$.

Let us now examine under which conditions the case of $V_2$ 
is recovered.  It turns out that
$$
\eqalign{
V_3(q_{12}=0,q_{13}=0,q_{23})=& V_2(q_{23})    \cr
V_3(q_{12}=0,q_{13},q_{23}=0)=& V_2(q_{13})\ . \cr}
\eq{caso22} 
$$
This is just what we expected for the following reason:
setting $q_{12}=0$ replica 2 thermalizes as in 
the free case and so replica 1 and 2  are both
equilibrium configurations. In this situation, if we set, for example,
$q_{13}=0$, the constraint 1-3 is harmless and the pair 2-3 reproduces
the case of $V_2$; the same is true for $q_{23}=0$.

\autosez{v3q} The shape of the potential.
\par

The most important information in the study of $V_3$ comes from the
analysis of its minima in the plane $(q_{13},q_{23})$ at fixed values
of $q_{12}$, which correspond to stable or metastable states of
the system.
We then minimize $V_3$ with respect of $q_{13}$ and $q_{23}$:
$$
\eqalign{
\beta^2 f'(q_{13})= &  {1\over 1-s_1+x_s(s_1-s_0)}
[q_{13}-q_{12}(q_{23}-w_{23})(a+b x_r)  \cr
           - & q_{12}(w_{23}-z_{23}) x_r (a+b x_r)]          \cr
\beta^2 f'(q_{23})= &  {1\over 1-s_1+x_s(s_1-s_0)}
[(q_{23}-w_{23})(a+b+c)      \cr
          + & (w_{23}-z_{23})(a+x_r(b+c))+(z_{23}-q_{12}q_{13})(a+b x_r)]  
                      \ .  \cr   }
\eq{min}
$$ 
First of all, we note  that for each value of $q_{12}$ there is
a solution of \re{min} for $q_{13}=q_{23}=0$, with  $V_3=0$.
This is the absolute minimum of the potential in the
$(q_{13},q_{23})$ plane and corresponds to the situation in which
the third replica sees a phase space identical to that of the 
unconstrained system 
\nota{
The requirement that  replica 3 is sited somewhere in 
the phase space, translates into the formula:
$\s \int dq_{13} dq_{23} e^{-N V_3(q_{13},q_{23})} = 1 \quad 
\forall q_{12} ,$
and for $\s N\rightarrow \infty$ this implies that in the absolute minimum
the potential has to be zero, as it is in our case.
}.

For what concerns non trivial minima, the numerical analysis of equations 
\re{min} in the case $p=3$, shows the following pattern:

i) For any value of $q_{12}$ and for any temperature in the range $(T_c,T_d)$,
we find a minimum $M_1$ with
$$
\eqalign{
q_{13}= & \ q_{EA}  \cr
q_{23}\sim & \  q_{12} \cr
V_3(M_1)= & \ V_2(q_{EA})  \ .  \cr}
\eq{m1}
$$
The interpretation of $M_1$ is very simple: the third replica is in the same
pure state of the first one, independently from the value of $q_{12}$; 
$M_1$ has exactly the same meaning of $M$ for the two replica potential. 
As $q_{12}$ grows, the second replica approaches this state, not changing the 
potential, until for $q_{12}=q_{EA}$ the three replicas are all in the same 
state and $q_{12}=q_{13}=q_{23}=q_{EA}$. We call this point $\Omega$.

It is important to note that $q_{23}=q_{12}$ holds only in three points (see
Fig. 2). Obviously it happens for $q_{12}=0$. Then, as $q_{12}$ grows,
we have $q_{12} > q_{23}$ and this holds until $q_{12}=q_{max}$, where
$q_{max}$ is the value for which $V_2$ has its maximum;
here again $q_{12}=q_{23}$. For $q_{max} < q_{12} < q_{EA}$ we have the 
opposite situation, with $q_{12} < q_{23}$, until we reach $\Omega$.  

This behaviour is consistent with the idea that at a distance $q_{max}$
from the state of replica 1 the free energy reaches a maximum; the 
interpretation of Fig. 2 is then the following:
Replicas 1 and 3 are in the same state but they have not the same role:
replica 1 is some quenched typical configuration of the state, 
while, in some sense, replica 3 represents the center of the state,
because of the thermodynamic average. When $q_{12}<q_{max}$, replica 2
reaches its minimum energy taking the maximum distance from the center 
of the state, compatibly with the given value of $q_{12}$, and in so
doing $q_{23}$ results lower than $q_{12}$. 
The opposite situation holds when $q_{12}>q_{max}$.

\vfill\eject

\includegraphics{snacio_fig2.ps}

\vbox{
      \hbox{     \vbox{$\s q_{23}$ \vglue 2.7 truecm}
                 \vbox{\vglue 6.5 truecm}                             }
      \hbox{     \hglue 5.5 truecm $\s q_{12}$                        }
      \hbox{     \piccolo{\hskip 1 truecm Fig. 2: $\s q_{23}$ in 
                 $\s M_1$ as a function of $\s q_{12}$ at 
                 $\s \beta=1.64$; $\s q_{max}=0.361$,
                 $\s q_{EA}=0.546$.}                                  }
                                                                            }

ii) More interesting is the presence of a minimum with the third replica
close to the second one.
For $T$ in the usual range $(T_c,T_d)$ and $0\leq q_{12} \leq \bar q (T)$
we find a minimum $M_2$ with
$$
\eqalign{
q_{23}\sim & \ q_{EA}   \cr
q_{13}\sim & \ q_{12} \leq  \ q_{EA} \ .  \cr}
\eq{m2}
$$
It is important to note that the last value $\bar q$ of $q_{12}$ for which 
$M_2$ exists is at any temperature less or equal to $q_{EA}$.
 
\includegraphics{snacio_fig3a.ps}

\vbox{
      \hbox{     \vbox{$\s q_{23}$ \vglue 2.7 truecm}
                 \vbox{\vglue 6.5 truecm}                             }
      \hbox{     \hglue 5.5 truecm $\s q_{12}$                        }
      \hbox{     \piccolo{\hskip 1 truecm Fig. 3a: $\s q_{23}$ in 
                 $\s M_2$ as a function of $\s q_{12}$ at 
                 $\s \beta=1.64$; $\s q_{EA}=0.546$, $\s \bar q=0.385$.}    }
                                                                            }
\eject

\includegraphics{snacio_fig3b.ps}

\vbox{
      \hbox{     \vbox{$\s \Delta E$ \vglue 2.7 truecm}
                 \vbox{\vglue 6.5 truecm}                             }
      \hbox{     \hglue 5.5 truecm $\s q_{12}$                        }
      \hbox{     \piccolo{\hskip 1 truecm Fig. 3b: The energy
                 difference $\s \Delta E$ in 
                 $\s M_2$ as a function of $\s q_{12}$ at 
                 $\s \beta=1.64$; $\s q^\star=0.295$, $\s \bar q=0.385$.}   }  

                                                                         }

\includegraphics{snacio_fig3c.ps}

\vbox{
      \hbox{     \vbox{$\s V_3$ \vglue 2.7 truecm}
                 \vbox{\vglue 6.5 truecm}                             }
      \hbox{     \hglue 5.5 truecm $\s q_{12}$                        }
      \hbox{     \piccolo{\hskip 1 truecm Fig. 3c: The potential 
                 $\s V_3$ in 
                 $\s M_2$ as a function of $\s q_{12}$ at 
                 $\s \beta=1.64$; $\s q^\star=0.295$, $\s \bar q=0.385$.}      }
                                                                            }
\noindent
Let us fix a reference temperature $T$ and examine $M_2$ at different
values of $q_{12}$. For $q_{12}=0$ we have $q_{13}=0$, $q_{23}=q_{EA}$ and 
$V_3(M_2)=V_2(q_{EA})$; this is trivial because we have seen that in
this case $V_3$ reduces to $V_2$.  When $q_{12}\neq 0$  we force the second
replica to thermalize in a restricted portion of the phase space; 
the fact that we still find the  minimum $M_2$ corresponding to the third 
replica in equilibrium close to the second one, shows that 
there are  local equilibrium states even at non 
zero overlap $q_{12}$ with the state of the first free replica. 
The variation with $q_{12}$ of the interesting quantities evaluated in $M_2$
is shown in Fig. 3.

We note that $q_{23}$ remains close to $q_{EA}$, confirming the 
hypothesis that the second and the third replica are in the same state.
It is then crucial to verify if the states corresponding to the minima
$M_2$ can be identified with TAP solutions, i.e. if their free
energy and self-overlap satisfy the TAP equations.
First, at fixed $q_{12}$, the self-overlap of the state relative to
$M_2$ is given by $s_1$ (see eq.\re{qq33}). Secondly, we must
compute the free energy of this particular state; to this end we can write
$$
V_3(M_2)=(f_3-T\Sigma_3) - F_{free}  \ ,
\eq{zut}
$$
where $f_3$ is the free energy of the state in which replica 3 thermalizes,
and $\Sigma_3$ is the logarithm of the number of states accessible to
replica 3, compatibly with the constraints; obviously, since replica 3
is in the same state of replica 2, which is fixed, we have $\Sigma_3=0$ and
therefore
$$
f_3=V_3(M_2)+F_{free}  \ .
\eq{f3}
$$
It turns out that in $M_2$ the relation holding between $f_3$ 
and $s_1$ is always the same of the TAP approach \cite{kpz}. In
other words, the state corresponding to $M_2$ coincides with that
particular TAP solution specified by
$$
\eqalign{
f_{TAP}=&\ f_3  \cr
q_{TAP}=&\ s_1 \ . \cr}
\eq{tap}
$$

Let us now consider the behaviour in $M_2$ of the 
energy difference $\Delta E=E-E_{free}$: we note that there is a value 
$q_{12}= q^\star$ for which
$$
\Delta E(q^\star)=\ 0 \ .
\eq{ener}
$$
The fact that in this minimum the energy is equal to the one of the 
unconstrained 
system means that replicas 2 and 3 are in a state of the same kind of 
the one chosen by the unconstrained system, that is one of the TAP solutions 
dominating the equilibrium; indeed, we have 
$s_1(q^\star)=q_{EA}$ (the same  happens in the minimum $M$ of 
$V_2$). For this reason, the valley entropic contribution $S_v$ to $V_3$  
is the same as in the free case (see eq.\re{uno}).
Thus we expect
$$
V_3(q^\star)=T\Sigma=V_2(q_{EA}) \ ,
\eq{gnucco}
$$ 
that is what we find (see Fig. 3c).

Therefore, we have the following picture: at temperature $T$ with
$T_c<T<T_d$, there is an exponentially high number of equilibrium states 
partitioning the phase space and the first unconstrained replica will thermalize
into one of them, call it $K$. From our discussion follows that the closest 
states of the same kind and with the same energy density of $K$, are found at 
overlap $q^\star$ with $K$. Moreover, the fact that the last minimum $M_2$
is found at $q_{12}=\bar q$, shows that at smaller distances there are local 
equilibrium states with higher energy density, the closest of which have 
overlap $\bar q$ with $K$.   

\autosez{v3T} Temperature dependence.
\par

Up to this moment we have studied $V_3$ at fixed temperature; now
we examine the evolution of the system with $T$, in particular in
the two limits $T\rightarrow T_c$ and $T\rightarrow T_d$.

At $T_c$ the configurational entropy $\Sigma$ of the dominant 
equilibrium states goes to zero \ccite{kpz}{crisatap}; this means
that the number of these states becomes of order $N$ and 
they {\it all} have zero overlap with each other. In this 
situation we expect that, fixed an equilibrium state, the closest 
one is at overlap zero; indeed, we find that $q^\star\rightarrow 0$ for
$T\rightarrow T_c$. On the other side, at $T_c$ the equilibrium
states have the lowest energy density \ccite{kpz}{crisatap}, while
there is an exponentially high number of local equilibrium
states with higher energy; according to this we find a non zero
value of $\bar q$ even at $T_c$.

What happens at $T_d$ is more interesting. First, we know that at
this temperature the equilibrium states have the highest energy
density, so we expect not to find local equilibrium states at
higher energy; this is exactly what we have, since 
$\bar q\rightarrow q^\star$ for $T\rightarrow T_d$.
Moreover, at this temperature  $\Sigma$ reaches a finite value, meaning that
the number of the equilibrium states is still exponentially high; 
therefore, it is not clear {\it a priori} what is their minimum
mutual distance. The very interesting feature shown by our potential
is that
$$
\bar q, q^\star \rightarrow q_d \quad{\rm for}\quad  T\rightarrow T_d \ ,
\eq{zundap}
$$
where $q_d$ is the value of $q_{EA}$ at the dynamical transition.
Equation \re{zundap} means that as $T$ approaches $T_d$, 
the closest equilibrium states collapse into a single state of 
self-overlap $q_d$. 

\noindent
More precisely, for $T\sim T_d$ and $p=3$, we find
$$
\eqalign{
q_{EA} =& \ (T_d-T)^{1/2} + q_d  \quad , \quad q_d=0.5     \cr
q^\star =& \ a\  (T_d-T)^b + c  \ .  \cr}
\eq{fitea}
$$
A fit of the numerical data gives
$$
\eqalign{
a& = -1.2   \cr
b& = 0.24   \cr
c& = 0.5004  \ . \cr }
\eq{valori}
$$
As we have seen, this indicates that $q^\star$ reaches $q_d$ at $T_d$.
The value of the exponent $b$ can be less firmly established; indeed,
if we make a fit fixing $c=0.5$ we get
$$
b=0.30  \ .
\eq{fika!}
$$

Finally, we can study the potential for $T>T_d$. In this range the
dominant equilibrium state is the paramagnetic one and the two
replica potential has just the minimum in zero; for this reason 
the minimum $M_1$ of $V_3$ disappears, and do not exist other
minima with $\Delta E=0$. However we still find the minimum $M_2$
in a restricted range of $q_{12}$ (see Fig. 4), to point out the existence 
of local equilibrium states even at temperature greater than $T_d$,
corresponding to TAP solutions surviving above $T_d$. 
This situation holds up to a temperature $T_{last}>T_d$ over which
all non trivial minima disappear. For $p=3$ we have
$$
\eqalign{
\beta_c=& \ 1.706  \cr
\beta_d=& \ 1.633  \cr
\beta_{last}=& \ 1.573  \ . \cr}
\eq{temp}
$$
With regard to this see also \cite{buribarrameza}.

\includegraphics{snacio_fig4.ps}

\vbox{
      \hbox{     \vbox{$\s \Delta E$ \vglue 2.7 truecm}
                 \vbox{\vglue 6.5 truecm}                             }
      \hbox{     \hglue 5.5 truecm $\s q_{12}$                        }
      \hbox{     \piccolo{\hskip 1 truecm Fig. 4: The energy
                 difference $\s \Delta E$ in 
                 $\s M_2$ as a function of $\s q_{12}$ at 
                 $\s \beta<\beta_d$.}                                  }  

                                                                         }

\autosez{C1} Open questions and hints on the dynamics.
\par

Let us summarize some of our findings which are relevant for studying the 
dynamics.  Near each equilibrium valley there are local equilibrium states with 
energy density greater that the equilibrium one.  These local equilibrium states 
do exist up to a value $\bar q$ of the overlap, while the nearest equilibrium
states are found at overlap $q^\star$. At the dynamical transition 
the three values $\bar q$, $q^\star$ and $q_d$ merge.

The most surprising consequence of this picture is that the equilibrium states 
are not uncorrelated, i.e. they are not distributed randomly on the sphere, which 
would be the simplest possibility. Indeed, if they were randomly distributed, 
the nearest ones would be always at an  overlap  $q$, given by
$$
\ln\left(1-{q^2\over q_{EA}^2(T)}\right)=- 2 \Sigma (T) \ ,
\eq{log}
$$
which should remain definitively different from $q_d$ when $T \to T_d$, since
$\Sigma(T_d)$ has a finite value.

It seems that the equilibrium valley, that at $T_d$ has a flat direction (the 
replicon eigenvalue is zero), bifurcates in a bunch of not too different 
valleys below $T_d$, like in the SK model.
However, the organization of equilibrium states is not the same as in the SK 
model, 
because it shows up only when we add a constraint to the system.  It may be 
possible, as suggested in \cite{vira},
that this situation may be in some way described by a non-monotonous function 
$q(x)$, as for example the one found in \ccite{kpz}{ferrero}.

We can now ask what happens to a configuration starting at time zero 
near an equilibrium one.  In particular we are interested to compute 
the typical behaviour of the overlap $q(t)$ among the configuration at time zero 
and the configuration at time $t$. It is natural to suppose that the form of the 
function 
$q(t)$ will be typical of a punctuated equilibrium: exponentially long period
of stasis, where $q(t)$ fluctuates around a value that corresponds to a local 
equilibrium state, punctuated with fast variations of $q(t)$, which correspond to 
jumps between one local equilibrium and other local equilibrium states.

According to the previous picture the following scenario is rather likely.
After a short transient time  $q(t)$ will go to a value equal to the 
self-overlap of the 
valley, i.e. $q_{EA}$. After some exponentially large time it will jump to one of 
the nearest equilibrium states. If the barriers increase with increasing the 
distance (and therefore with decreasing the overlap) the most likely  situation 
consists of a jumping to a local equilibrium state with overlap $\bar q$.

What happens at later times is not clear. If the different local equilibrium 
states are not correlated among themselves, (many replica potential is needed 
to investigate this point further), the system will return to the original 
state after some exponentially large time. Only after many attempts the system 
will jump to another equilibrium state at distance $q^\star$. At this point the 
system will not likely jump back and it will start from this point to do further 
jumps.  

The scenario we have presented here seems to be the simplest one compatible with 
our finding on the three replica potential. In a future paper we plan to 
compare this proposal with numerical simulations. It would also be interesting 
to study the problem in the Random Energy Model, where many detailed 
computations can be done. 


\vskip 1 truecm
\noindent
{\bf Acknowledgments.}

It is a pleasure to thank Silvio Franz for very useful suggestions and 
discussion.

\vskip 3 truecm
\semiautosez{\rm A} Appendix A.
\par

In this appendix we calculate explicitly the final form of the 
potential $V_2(q_{12})$.
We have:
$$
-\beta V_2(q_{12})=
\lim_{n,m\rightarrow 0} \ {1\over 2m}
\left( \beta^2 \sum_{a,b} f({\bf Q}_{ab})+
\log \det {\bf Q}\right)  + \beta F_{free}   \ .
\eq{asnaz2}
$$
This expression has to be computed according to the particular form 
assumed for the overlap matrix.
The first part is
$$
\lim_{n,m\rightarrow 0} {1\over m}\beta^2\sum_{ab} f({\bf Q}_{ab})=
$$
$$
2\beta^2 f(q_{12})+
\beta^2 f(1)+\beta^2 (x_r-1)f(r_1)-\beta^2 x_r f(r_0)  \ .
\eq{asemp}
$$
The second part is the most complicated one. It is useful to exploit the 
following relation: 
$$
\log \det {\bf Q}= \log {\rm det} Q^{11} +
\log {\rm det} \left( Q^{22}- Q^{21} (Q^{11})^{-1} Q^{12} \right) \ ,
\eq{acomp}
$$
where   $Q^{21}$ stands for $(Q^{12})^T$.
With our ansatz, from \re{q11} and \re{q12}
$$
\log {\rm det} Q^{11}=0 \ ,
\eq{azz}
$$ 
and
$$
Q^{21}(Q^{11})^{-1}Q^{12}=q_{12}^2 \ ,
\eq{azz1}
$$
is a constant matrix. In this way equation \re{acomp} becomes
$$
\log \det {\bf Q}= \log {\rm det}R \ ,
\eq{adetR}
$$ 
with
$$
R_{ab}=Q^{22}_{ab}-q_{12}^2=(1-r_1)\delta_{ab}+(r_1-r_0)\epsilon_{ab}
+(r_0-q_{12}^2)  \ .
\eq{aerre}
$$
To calculate the determinant of $R$ we need to solve the eigenvalue equation
$$
(1-r_1)v_a+(r_1-r_0)\sum_b\epsilon_{ab}v_b+(r_0-q_{12}^2)\sum_b v_b=
\lambda v_a  \ .
\eq{aeqaut}
$$
Let us distinguish three cases:

{\noindent i)}\ If 
$$
 \sum_b\epsilon_{ab}v_b=0 \quad \quad \forall \quad a \ ,
\eq{acaso1}
$$
and, consequently,
$$
\sum_b v_b=0 \ ,
\eq{acaso1.2}
$$
we have
$$
\lambda_1=(1-r_1)  \ .
\eq{aaut1}
$$
In \re{acaso1} the number of independent 
equations is equal to the number of blocks, that is $m/x_r$; the 
degeneration of the previous eigenvalue is then
$$
d_1=m-m/x_r \ .
\eq{adeg1}
$$

{\noindent  ii)}\ If 
$$
\sum_b v_b=0 \ ,
\eq{acaso2}
$$
we obtain:
$$
\lambda_2=(1-r_1)+x_r(r_1-r_0)  \  ,
\eq{aaut2}
$$
because the elements of the vector $\bf v$ must be equal in blocks of size $x_r$.
The number of equations determining $\lambda_2$ is one and thus we have
$$
d_2=m-(m-m/x_r)-1=m/x_r-1 \ .
\eq{adeg2}
$$

{\noindent iii)} \ If the previous conditions are not satisfied, 
and the sums are all different from zero equation \re{aeqaut} gives 
$$
\lambda_3=(1-r_1)+x_r(r_1-r_0)+m(r_0-q_{12}^2)  \ ,
\eq{aaut3}
$$
$$
d_3=1 \ ,
\eq{adeg3}
$$
Expression \re{adetR} then becomes
$$
\eqalign{
{1\over m} \log {\rm det} R= \log (1-r_1)+ & {1\over x_r}\log\left(
1+x_r{r_1-r_0\over 1-r_1}\right)                \cr
                                         + &{1\over m}\log\left(
1+m{r_0-q_{12}^2\over 1-r_1+x_r(r_1-r_0)}\right)
\bbuildrel\hbox to .4in{\rightarrowfill}_{m\rightarrow 0}  \cr
\log(1-r_1)+ & {1\over x_r}\log\left(
1+x_r{r_1-r_0\over 1-r_1}\right)+
{r_0-q_{12}^2\over 1-r_1+x_r(r_1-r_0)}  \  .    \cr}
\eq{adetfin}
$$
From \re{asemp} and \re{adetfin} we get the final form 
$$
\eqalign{
-2\beta V_2(q_{12})= &               
\ 2\beta^2 f(q_{12})+
\beta^2 (x_r-1)f(r_1)-\beta^2 x_r f(r_0)     \cr
+ & \log(1-r_1)+{1\over x_r}\log\left(
1+x_r{r_1-r_0\over 1-r_1}\right)+
{r_0-q_{12}^2\over 1-r_1+x_r(r_1-r_0)}    \  . \cr}       
\eq{apotfin}
$$

For the sake of completeness we note that there is a different method
by  which computing the potential, based on the formula
$$
- \beta N V_2(q_{12})=
$$
$$
\eqalign{
 =\lim_{n \rightarrow 0}\lim_{R \rightarrow 1}{1 \over n} &
\overline{
{\partial\over\partial R} \left( \int d\sigma \exp (-\beta H(\sigma))
\left( \int d\tau \exp (-\beta H(\tau))  \delta (q_{\sigma\tau}-q_{12})
\right)^{R-1}\right)^n   }                                 \cr
& \phantom{{\partial\over\partial R} \left( \int d\sigma \exp (-\beta H(\sigma))
\left( \int d\tau \exp (-\beta H(\tau))\right)\right)   }
-\overline{\log Z_{free}}   \ ,                            \cr}
\eq{replica2}
$$
where in the calculation of the average $R$ has to be considered as an integer 
(compare with eq. \re{replica}).
In this way the replicated partition function takes the form
$$
Z^{n,R}=
\overline{
\int d{\tau}_a^r \exp \left[- \beta \left (\sum_{a=1}^n \sum_{r=1}^R
H(\tau_a^r)\right)\right] \prod_{a=1}^n\prod_{r=2}^R
\delta (\tau_a^1\tau_a^r-q_{12})  }  \ ,
\eq{rep3}
$$
where we set $\sigma_a=\tau_a^1$.
With this approach we obtain a $nR\times nR$  global overlap matrix $\bf Q$,
formed by the $n\times n$ sub-matrices $Q_{ab}^{rs}$ with $a,b=1,\dots ,n;\ 
r,s=1,\dots ,R$. In this way, the matrix $Q^{11}$ encodes the overlap
1-1, the matrices $Q^{1r}$ the overlap 1-2 and the matrices $Q^{rs}$ with
$r,s\neq 1$ encode the overlap 2-2. 
The ansatz that has to be taken for the various matrices is a mere 
translation in this context of the one exposed in Sect. 2.

The same line of reasoning can be applied in the computation of the
three replica potential. Obvioulsy, in both cases, the two methods are
equivalent.

\semiautosez{\rm B} Appendix B.
\par

In this appendix we calculate the final form of the potential $V_3$.
We have:
$$
-\beta V_3(q_{12},q_{13},q_{23})=
\lim_{n,m,l\rightarrow 0}\ {1\over 2l} 
\left( \beta^2 \sum_{a,b} f({\bf Q}_{ab})+
\log \det {\bf Q} \right) + \beta  F_{free} \ .
\eq{btuka}
$$
The first part is
$$
\eqalign{
\lim_{n,m,l\rightarrow 0} {1\over l} \beta^2 \sum_{ab} f({\bf Q}_{ab})= &
2\beta^2 f(q_{13})+2\beta^2 f(q_{23})+2\beta^2(x_r-1) f(w_{23})-
2\beta^2 x_r f(z_{23})      \cr
& +\beta^2 f(1)+\beta^2 (x_s-1) f(s_1)-\beta^2 x_s f(s_0)  \ .\cr}
\eq{bsemp}
$$
For the second part we use the following relation:
$$
\eqalign{
& \log\det{\bf Q}=                                 \cr
& = \log\det Q^{11}                                \cr
& + \log\det[Q^{22}-Q^{21}(Q^{11})^{-1}Q^{12}]     \cr
& + \log\det[Q^{33}-Q^{31}(Q^{11})^{-1}Q^{13}      \cr 
&\phantom{++}-(Q^{23}-Q^{21}(Q^{11})^{-1}Q^{13})^T
(Q^{22}-Q^{21}(Q^{11})^{-1}Q^{12})^{-1}
(Q^{23}-Q^{21}(Q^{11})^{-1}Q^{13})]        \ ,                   \cr}
\eq{bpisto}
$$
where we adopted the previous notation for the transposed matrices.
The first two parts in the r.h.s reproduce the case of the two replicas
potential and both go to zero when $(n,m,l)\rightarrow 0$.

The remaining parts of the determinant are:
 
$$
\eqalign{
Q^{31}(Q^{11})^{-1}Q^{13}=& \ q_{13}^2    \cr
Q^{21}(Q^{11})^{-1}Q^{13}=& \ q_{12}q_{13}   \cr
Q^{21}(Q^{11})^{-1}Q^{12}=& \ q_{12}^2  \ ,  \cr}
\eq{bd}
$$
constant matrices. Finally, let us compute the inverse of $A=Q^{22}-q_{12}^2$:
$$
A_{ab}=(1-r_1)\delta_{ab}+(r_1-r_0)\epsilon_{ab}+(r_0-q_{12}^2)  \  ,
\eq{ba}
$$
that is,
$$
(A^{-1})_{ab}={1\over 1-r_1}\ \delta_{ab}+
{r_0-r_1\over 1-r_1}\cdot {1\over 1-r_1+x_r(r_1-r_0)}\ \epsilon_{ab}+
{q_{12}^2-r_0\over (1-r_1+x_r(r_1-r_0))^2} \ .
\eq{bsuka}
$$
Now it is possible to calculate the product 
$$
\pi=
(Q^{23}-Q^{21}(Q^{11})^{-1}Q^{13})^T
(Q^{22}-Q^{21}(Q^{11})^{-1}Q^{12})^{-1}
(Q^{23}-Q^{21}(Q^{11})^{-1}Q^{13})  \ .
\eq{bpru}
$$
As can be easily seen the three matrices have the correct form, since they are 
RSB matrices with the same breaking point $x_r$. Due to this fact, the result
of this product is a constant, although very complicate, given in eq. \re{y2}.

Now it remains to calculate:
$$
\log\det[Q^{33}-y]   \ ;
\eq{blog}
$$
proceeding as in Appendix A, with $y$ instead of $q_{12}^2$, we get 
$$
{1\over l}\log\det[Q^{33}-y]=
\log(1-s_1)+{1\over x_s}\log\left(
1+x_s{s_1-s_0\over 1-s_1}\right)+
{s_0-y\over 1-s_1+x_s(s_1-s_0)}   \  .
\eq{bfin}
$$
In this way we recover equation \re{v3fin}.

\vfill\eject
\vskip 2 truecm
\noindent {\bf References.}
\vskip 0.5 truecm



\biblitem{ea} S.F. Edwards, P.W. Anderson, {\it J. Phys.} F {\bf 5} (1975), 
965.

\biblitem{sk} D. Sherrington, S. Kirkpatrick, {\it Phys. Rev. Lett.} {\bf 35}
(1975), 1792.

\biblitem{tap} D.J. Thouless, P.W. Anderson, R.G. Palmer, {\it Philos.
Mag.} {\bf 35} (1977), 593.

\biblitem{rsb1} G. Parisi, {\it Phys. Rev. Lett.} {\bf 23} (1979), 1754; 
                          
\biblitem{rsb2} G. Parisi, {\it J. Phys.} A {\bf 13} (1980), L115; 
                                           
\biblitem{rsb3} G. Parisi, {\it J. Phys.} A {\bf 13} (1980), 1887.

\biblitem{sompozip} H. Sompolinsky, A. Zippelius, {\it Phys. Rev.} B {\bf 25}
(1982), 6860.

\biblitem{ck1} L.F. Cugliandolo, J. Kurchan, {\it Phys. Rev. Lett.}
{\bf 71} (1993), 173.

\biblitem{tirumma} T.R. Kirkpatrick, D. Thirumalai, {\it Phis. Rev.} B {\bf 36} 
(1987), 5388.

\biblitem{crisa1} A. Crisanti, H.J. Sommers, {\it Z. Phys.} B {\bf 87} (1992), 
341.

\biblitem{crisa2} A. Crisanti, H. Horner, H.J. Sommers, {\it Z. Phys.} B
{\bf 92} (1993), 257.

\biblitem{ck2} L.F. Cugliandolo, J. Kurchan, {\it J. Phys.} A {\bf 27} 
(1994), 5749.

\biblitem{kpz} J. Kurchan, G. Parisi, M.A. Virasoro, {\it J. Phys. I France}
 {\bf 3} (1993), 1819.

\biblitem{franzparisi}  S. Franz, G. Parisi, {\it J. Phys. I France}
{\bf 5} (1995), 1401.

\biblitem{ferrero} M.E. Ferrero, M.A. Virasoro, {\it J. Phys. I France }
{\bf 4} (1994), 1819.

\biblitem{buribarrameza} A. Barrat, R. Burioni, M. M\'ezard, {\it J. Phys} A 
{\bf 29} (1996), L81.

\biblitem{monasson} R. Monasson, {\it Phys. Rev. Lett.}
{\bf 75} (1995), 2847.

\biblitem{crisatap} A. Crisanti, H.J. Sommers, {\it J. Phys. I France}
{\bf 5} (1995), 805.

\biblitem{mezpa} M. M\'ezard, G. Parisi, {\it J. Phys. A }
{\bf 23} (1990), L1229; {\it J. Phys. I France }
{\bf 1} (1991), 809.

\biblitem{nieu} Th.M. Nieuwenhuizen, {\it Phys. Rev. Lett. }
{\bf 74} (1996), 4289.

\biblitem{franz} S. Franz, private communication.

\biblitem{vira} M.A.Virasoro, {\sl Simulated annealing methods under analytical 
control}, in Pro\-cee\-dings of 19th Intl. Conf. on Stat. Phys. (IUPAP),
Xiamen, China; to be published.

\biblitem{grome} D.J. Gross, M. M\'ezard, {\it Nucl. Phys. B }
{\bf 240} (1984), 431.

\biblitem{gard} E. Gardner, {\it Nucl. Phys. B}
{\bf 257} (1985), 747.

\insertbibliografia

\bye